# Real-time estimation of the short-run impact of COVID-19 on economic activity using electricity market data

Carlo Fezzi[1,2] and Valeria Fanghella[1]

June 2020

**Abstract:** The COVID-19 pandemic has caused more than 8 million confirmed cases and 500,000 death to date. In response to this emergency, many countries have introduced a series of social-distancing measures including lockdowns and businesses' temporary shutdowns, in an attempt to curb the spread of the infection. Accordingly, the pandemic has been generating unprecedent disruption on practically every aspect of society. This paper demonstrates that high-frequency electricity market data can be used to estimate the causal, short-run impact of COVID-19 on the economy. In the current uncertain economic conditions, timeliness is essential. Unlike official statistics, which are published with a delay of a few months, with our approach one can monitor virtually every day the impact of the containment policies, the extent of the recession and measure whether the monetary and fiscal stimuli introduced to address the crisis are being effective. We illustrate our methodology on daily data for the Italian day-ahead power market. Not surprisingly, we find that the containment measures caused a significant reduction in economic activities and that the GDP at the end of in May 2020 is still about 11% lower that what it would have been without the outbreak.

**Keywords**: COVID-19, coronavirus, economic impacts, lockdown, GDP, electricity quantity, wholesale electricity markets, pandemic, fixed-effect regression, high-frequency estimates, real-time monitoring.

---

[1] Department of Economics and Management, University of Trento, Via Vigilio Inama 5, 38122 Trento, Italy. Corresponding Author: Carlo Fezzi, email: carlo.fezzi@unitn.it.
[2] Land, Environment, Economics and Policy Institute, Department of Economics, University of Exeter Business School, Exeter, UK.

**Introduction**

COVID-19, i.e. the disease caused by Severe Acute Respiratory Syndrome Coronavirus 2, SARS-COV-2, first struck in China's Hubei Province at the end of year 2019 (Zhu et al., 2020). It quickly spread across the globe and was recognized by the World Health Organization (WHO) as a pandemic on the 11th of March 2020 (WHO, 2020), due to its high contagiousness and aggressive course. Thus far, it caused more than 8 million confirmed cases and about 500,000 deaths. In order to reduce the pace of the infection, most countries introduced a variety of containment strategies such as lockdowns, international and national traveling restrictions, social-distancing and shutdowns of non-essential businesses, schools and public offices. These measures have been saving lives by reducing the contagion and by alleviating the burden on health care systems (e.g. Anderson et al., 2020). However, they have also been generating substantial disruption on practically every aspect of society.

Preliminary studies indicate that the reduction in economic activity has been extremely significant. According to the most recent provisional estimates (Eurostat, 2020; Organization for Cooperation and Development, OECD, 2020), practically all developed economies are experiencing a contraction in the first quarter of year 2020. Italy and France are amongst the most affected countries and their Gross Domestic Products (GDPs) are estimated to have fall by 5.3%. This is a very significant downturn. As a comparison, the GDP of these two economies decreased respectively by 6.3% and 2.6% during the entire financial crisis of the years 2008-09, aptly named "The Great Recession".[3]

Albeit informative, official statistics have two main drawbacks. First, they are typically released with a delay of about three months.[4] Second, they simply provide a picture of the status of the economy, but cannot disentangle the impact of COVID-19 from those of all other factors affecting production and consumption. In other words, they do not offer any causal estimates of the impact that the virus and the related containment measures are having on economic activities. However, in this time of uncertainty and economic downturn, policy makers are in strong need of timely indicators to *a)* monitor in real-time the evolution of the economy and *b)* understand the causal impact of the policies designed to respond to the COVID-19 pandemic, including both those implemented to contain the virus and those developed to stimulate production and consumption after restrictions are lifted.

---

[3] Source: The World Bank, https://data.worldbank.org/.
[4] As an example, economic assessments for the fourth quarter of 2019 were available only in April 2020 for both the European Union (Eurostat, 2020b) and the United States (Bureau of Economic Analysis, BEA, 2020).

This paper shows how high-frequency electricity market data can provide both types of information. Electricity is traded on hourly (or even half-hourly) bases in most developed countries across the world and up-to-date information on consumption is publicly accessible via the system operators' websites. In developed countries, electricity contributes to virtually every human activity. The relationship between electricity consumption (often indicated by the term "load") and economic development is well established in the energy economics literature (e.g. Kraft and Kraft, 1978; Chen et al., 2007; Stern, 2018). By taking advantage of this strong link, satellite data on night lights are used to provide geographically disaggregated GDP estimates (Henderson et al., 2012) and even to forecast economic growth (Galimberti, 2020). While some may argue that in a few developed countries this historical long-run relation is shifting because of the tertiarization of the economy (e.g. Buera and Kaboski, 2012) and the investments in energy-saving technologies, nobody would dispute that the short-run association between electricity load and economic activities is now as strong as ever. For example, significant drops in load occur during nighttime, weekends and public holidays when many businesses are shut down, creating the characteristic multi-level (daily, weekly and annual) seasonality of electricity markets' time-series (Bunn and Farmer, 1985). This paper develops a methodology to harness the wealth of information contained in electricity market data in order to monitor in real-time the state of the economy and to estimate the causal impact of the pandemic.

We illustrate our approach using daily data from the Italian day-ahead power market. Italy was the first European country to experience the outbreak of COVID-19 and, therefore, it is the one with the longest history of coping with the virus. It was also the first European country to implement a series of nation-wide restrictions on social and economic activities in an attempt to reduce the spread of the pandemic. Since the Italian Government introduced (and then lifted) a variety of measures of gradually increasing severity, our daily information allows us to separately estimate the short-run economic impact of each of the different restrictions. For all these reasons, Italy is a very informative case-study on both the impact of the containment strategies and on the possible path of economic recovery. Taking into consideration the appropriate caveats, our estimates provide also valuable insights on how economic activities may react to future restrictions.

Our identification strategy relies on a fixed-effect model and on daily electricity load information for the years 2015-20. We estimate the impact of COVID-19 on consumption and then apply a simple and yet effective rescaling approach in order to derive the implications for the GDP. Our best estimate of the impact of the virus in the first quarter of year 2020 is a 5.1% GDP reduction (with 95% confidence interval being between -4.0% and -6.2%), which is consistent with the official figures mentioned earlier. Of course, our data is considerably more chronologically disaggregated and up to date, so we are not limited to the first quarter. On the contrary, we can distinguish the dynamic impacts of the different policies and monitor the recovery of the economy thus far. We find that each newly introduced restrictive measure is followed by a significant drop in consumption. Not surprisingly, the strictest policies cause the largest reductions. For example, we estimate that the three weeks of most intense lockdown (with the temporary shut-down of a large number of factories) curtailed the GDP of that period by about 30%. In similar fashion, when Italy started the gradual resumption of economic and social activities, our estimated impact gradually diminishes. However, load has not yet reached the level that it would have been according to our counterfactual, signaling that the adverse impact of COVID-19 on the economy is still significant and ongoing.

**Electricity market data**

The Italian electricity market opened in year 2004, which makes it one of the youngest power markets in Europe. Most of Italian electricity consumption pertains to three large sectors: industry (39.6%), commercial and public services (32%) and residential (22.4%). The remaining 5.9% is allocated to transport, agriculture, forestry and fishing. The overall annual demand of about 300TWh is met by a mix of fossil fuels (about 65%), renewables (21%) and imports (14%).[5]

We calculate daily electricity load from the hourly, publicity-available information provided by the Italian day-ahead electricity market system operator, *Gestore dei Mercati Energetici* (GME).[6] Our dataset ranges from the 1$^{st}$ of January 2015 to the 31$^{st}$ of May 2020. To illustrate the key dynamics of electricity consumption and to provide a first, visual inspection of the impact of COVID-19, Figure 1 compares the daily load for years 2019 and 2020. We first consider year 2019, represented by the gray line. The weekly seasonality is pronounced, with the reduced business activity in the weekends translating into roughly a 20% drop in load. Similar sharp falls also characterize public holidays. While these features are common to all electricity markets across the world, in Italy the effect of economic activities on load is also visible from the substantial reduction in consumption during the two central weeks of August, which is when the majority of Italian businesses shut down for the summer break. We also observe a smoother, annual seasonality, which follows the path of temperature, with peaks during winter and summer, when consumption for air-conditioning is at the highest.

[ Figure 1 about here ]

---

[5] Source: International Energy Agency: https://www.iea.org/data-and-statistics.
[6] GME historical data is available at: https://www.mercatoelettrico.org/En/Tools/Accessodati.aspx.

The electricity quantity in year 2020 follows virtually the same path of year 2019 until the first week of March, when it starkly diverges. That is when the Italian Government started introducing a series of social-distancing measures in order to contain the escalating numbers of COVID-19 cases. On the 10$^{th}$ of March it was implemented the first national-level lockdown, which was quickly followed by the provisional closure of all shops, including bars and restaurants.[7] The hardest measures were introduced in the last week of March and were enforced until the second week of April. During this period, most factories were shut down and only supermarkets, pharmacies and other essential services were allowed to operate, but with reduced opening hours.[8] Not surprisingly, this is when the gap between the electricity loads of the two years is at the widest. Production and consumption steadily recovered in the following weeks, with the gradual re-opening of the economy. However, at the end of May we can still observe a substantial gap between the two years (the detailed timeline is reported in Table A1 in the Appendix).

[Figure 2 about here]

We examine the impact of COVID-19 containment policies more in detail in Figure 2, which compares the weekly seasonality across all years in our sample distinguishing between two periods: before the lockdown (weeks 5 to 9 in February and March) and during the lockdown (weeks 12 to 16 in March and April). On one hand, the pattern of electricity consumption in year 2020 before the lockdown appears to be close to the average level of the previous five years, arguably a result of the slow GDP growth and of increased energy-efficiency (e.g. Malinauskaite et al., 2019). On the other hand, during the lockdown weeks the difference between year 2020 and the previous years is remarkable. In that period, consumption in the weekdays of 2020 is comparable to that of the weekends of 2015-19, while during the weekends of 2020 we observe an even lower load.

All these findings confirm once more the robust short-run relationship between the intensity of economic activities and electricity consumption. As mentioned earlier, another important determinant of load is air temperature. To represent this variable, we use the average between the mean daily temperature in the two largest Italian cities: Rome and Milan. We download this information up to the 13$^{th}$ of May 2020 from the University of Dayton archive[9] and the observations for the remaining

---

[7] DPCM 9$^{th}$ of March, Official Gazette: https://www.gazzettaufficiale.it/eli/id/2020/03/09/20A01558/sg.
[8] DPCM 22$^{nd}$ of March, Official Gazette: https://www.gazzettaufficiale.it/eli/id/2020/03/22/20A01807/sg and DPCM 25$^{th}$ of March, Official Gazette: https://www.gazzettaufficiale.it/eli/id/2020/03/25/20G00035/sg.
[9] Source: http://academic.udayton.edu/kissock/http/Weather/.

days from the monitoring stations of the National Oceanic and Atmospheric Administration (NOAA).[10]

**Modelling approach**

The visual inspection of the data in the previous section highlighted that, up to the day in which the first containment measures were introduced, electricity consumption in year 2020 exhibited very similar patterns to those in the previous five years. Therefore, the information in those earlier years can be used to construct a plausible counterfactual for what electricity consumption would have been in year 2020 in the absence of the pandemic. This line of reasoning is the same one used to show that, in most countries, official figures greatly underestimate the real death toll of the outbreak (e.g., Ciminelli and Garcia-Mandicó, 2020).

Indicating with $t$ the daily steps of our time-series, our base model is:

(1) $y_t = \beta_0 + \sum_{w=1}^{6} \beta_w d_{wt} + \sum_{h=1}^{2} \beta_h d_{ht} + \gamma_w + \gamma^*_{w,2020} + f(temp_t) + u_t$ ,

where $y_t$ is the natural logarithm of electricity load, $d_{wt}$ are six dummy variables identifying the day of the week (with Monday as the baseline), $d_{ht}$ are two dummy variables identifying official public holidays and other observances, $\gamma_w$ = are week-of-the-year fixed effects, $\gamma^*_{w,2020}$ are week-of-the-year fixed effects interacted with a dummy variable identifying year 2020, $temp_t$ = is the average air temperature and $f(.)$ a non-linear functional form, $\beta$s are the remaining a parameters to be estimated and $u_t$ is the random component.[11]

This specification is designed to capture all the peculiar features of electricity consumption and to isolate the causal impact of COVID-19. The week-of-the-year fixed effects $\gamma_w$ encompass the slow-moving yearly seasonality of electricity load connected to both weather and cultural habits, such as the distinctive drop in consumption in the central two weeks of August that we presented in the previous section. The non-linear effect of the short-term variation in temperature is represented by $f(.)$, which we specify as joint piecewise linear function as $\delta_1 temp_t + \delta_2(temp_t - k)d_{kt}$, where $k$ is the threshold in which the relationship between temperature and load reverts and $\delta_1$, $\delta_2$ are the parameters to be estimated. This flexible specification generates a V-shaped function. We set $k$ to 62°F (about 16.5°C) by visually inspecting the scatter-plot of the data (see Figure A1 in the Appendix). Furthermore, the effects of the weekly seasonality and the public holidays are captured by the corresponding dummy variables.

---

[10] Source: https://www.ncdc.noaa.gov/cdo-web/
[11] Regarding the week-of-the-year fixed effects, we tested two alternative specifications. In the first one we used calendar weeks. With this approach, in different years the days covered by each fixed-effects do not correspond and the number of days represented by the first week can vary across years. In the second approach, we defined the fixed effects as groups of seven days, always starting from the 1st of January of each year. Therefore, the only difference across years is caused by leap years. This second method provided a slightly superior fit and, therefore, is the one we adopt in the all our models.

Our model does not include electricity price. In fact, a peculiar characteristic of electricity markets is that the demand function can be considered as completely inelastic in the short-run, since the majority of final consumers do not purchase on the wholesale exchange but, rather, are supplied by utility companies at fixed tariffs (e.g., Fezzi and Bunn, 2010). These companies do operate on the day-ahead market, but are required to fulfill their orders to the final consumers and, therefore, cannot respond to any price variation. Because of this feature, short-run electricity load forecasting models do not typically include any price information (e.g. Taylor et al., 2006) and, practically, all short-run price forecasting methods treat quantity as exogenous (e.g. Weron, 2014; Fezzi and Mosetti, 2020). In line with this long-standing literature, therefore, we exclude any price effect from our short-run analysis of electricity consumption.

The key-parameters for our study are the coefficients $\gamma^*_{w,2020}$, which measure the impact of COVID-19. These interaction effects capture the differences between each week of year 2020 and the average of the corresponding week in the previous five years which cannot be explained by any of the other observed factors. If our model is correctly specified the $\gamma^*_{w,2020}$ parameters corresponding to the weeks before the outbreak (i.e. during January and February) should be not significantly different from zero. Of course, we still include these parameters in our model because they serve as implicit time-placebo tests for our modelling assumptions. On the other contrary, we expect to estimate highly significant and negative $\gamma^*_{w,2020}$ parameters when the COVID-19 containment measures are introduced, i.e. roughly from the second week of March.

If the error component $u_t$ is independently and identically distributed (*iid*), our model can be consistently estimated with ordinary least squares (OLS). A possible concern with this estimator is that the *iid* assumption may not be satisfied, since the unobserved factors represented by the stochastic component may be autocorrelated. This situation can be generated by autocorrelated measurement errors. For example, the average temperature in Milan and Rome is unlikely to perfectly represent the weather profile of the entire country and, therefore, some remaining demand variation is undoubtedly present in the error term. Since weather shocks are typically autocorrelated, this missing variation is likely to generate autocorrelation in the residuals of our model. Possible omitted variables (e.g. special events, dynamic adjustments in residential and commercial load to temperature variations) are also plausible causes of residual autocorrelation. We address this issue following two distinct approaches. In the first one, we apply to the OLS covariance matrix the heteroscedasticity and autocorrelation consistent (HAC) correction proposed by Newey and West (1987). By setting the maximum lag for the correlation weights at seven we also attempt to capture any remaining weekly seasonality. In the second approach, we impose an autoregressive AR(1) specification for the random component (i.e. $u_t = \phi u_{t-1}$) and estimate the resulting model with maximum likelihood. As a further check of robustness of our findings, we re-estimate the base model after removing the piecewise function of temperature from the equation, in order to evaluate the susceptibly of our estimates to omitted variable bias. As shown in the next section, none of these alternative specifications changes our estimates of the effect of COVID-19 in any significant way. We run all our analysis in *R* (R Development Core Team, 2006), using the packages *lmtest* (Hothorn et al., 2019), *MASS* (Ripley et al, 2013), *nlme* (Pinheiro et al., 2017) and s*andwich* (Zeileis 2004).

If our model is correctly specified and passes the time-placebo test above, the coefficients $\gamma^*_{w,2020}$ from week 11 to week 22 can be interpreted as the causal impact of COVID-19 on electricity load. Therefore, we can estimate the impact of the pandemic by comparing daily in-sample predictions obtained by *a*) the full model and *b*) a model in which such coefficients are set to zero. Indicating these two predictions (on the original scale of the variable) respectively with $\hat{Y}_t$ and $\hat{Y}^*_t$, we can write the percentage impact of COVID-19 on load as:

(2) $l_t = 100(\hat{Y}_t - \hat{Y}^*_t)/\hat{Y}^*_t$,

and derive appropriate confidence intervals via Monte Carlo simulations.[12]

Ideally, in order to translate electricity load reductions into GDP impacts we would want to employ detailed information, disaggregated by industry type, on changes in electricity consumption, value added and amenability to distance-working solutions. Unsurprisingly, this wealth of data is not available, particularly at the daily time-scale of our analysis. As a second-best solution, we employ some deliberately simple and intuitive assumptions to transform our estimates of electricity load changes into GDP impacts.

---

[12] Because of Jensen's inequality (e.g. Silva and Tenreyro, 2006), the predictions on the original variable's scale are not simply the exponent of the predictions on the log, but actually depend on the distribution of the random component. The practical significance of this inequality is influenced by the magnitude of the residuals, i.e. by the noise-vs-signal ratio. In our specific case the difference is negligible since all our specifications present extremely high fit (all $R^2$ are higher than 0.92) and, therefore, for simplicity we employ the Gaussian assumption, i.e. we use $Y_t = \exp\{[\log(y_t)] + (s^2/2)\}$, with *s* indicating the estimated standard deviation of the error component.

We assume that, in the short run and at the national level, GDP changes are proportional to the changes in electricity consumption by all productive sectors (i.e. all sectors but the residential one). In order to evaluate this claim, we run a back-of-the-envelope calculation on GDP and electricity consumption information for Italy for the years 1990 to 2018.[13] Both variables are non-stationary and, therefore, we compute a correlation analysis on the first differences in order to avoid measuring a spurious relation. We estimate a correlation coefficient of 0.9, indicating an extremely strong linear covariation in the long-run. We believe this relationship to be even stronger in the short-run, justifying the simple assumption of a 1:1 relationship in our calculations.

In order to derive the impact of COVID-19 on the electricity consumption of the productive sectors we need to rescale our estimates, which are calculated on the total load. We compare two simple approaches. In the first one, we assume that residential consumption has remained unaffected by the restrictions and, therefore, all the reduction in electricity load due to the COVID-19 can be traced back to the other sectors. In the second one, we follow recent International Energy Agency's estimates (IEA, 2020) reporting that domestic consumption has increased by 40% during the lockdown, and rescale our calculations accordingly. The percentage GDP impacts following these two methods can be written as:

(3) $GDP_{1t} = l_t \, 100 / (100 - r)$, and
(4) $GDP_{2t} = l_t \, 100 / (100 - 1.4r)$,

where $r$ represents the percentage of consumption of the residential sector, which in Italy corresponds to 22.4% according to the most up-to-date IEA estimates.[3] While for transparency we report both measures, we believe that (4) should be the preferred estimate during the lockdown months (March and April) while (3) should be more accurate for the post-lockdown period, i.e. from the month of May onward. Although extremely simple, the next section shows that these assumptions provide results that are remarkably close to the official estimates of the GDP changes during the first quarter of 2020. Of course, our results extend well beyond the first quarter and provide an up-to-date assessment of the status of the economic disruption caused by the pandemic.

---

[13] Source: IEA, https://www.iea.org/data-and-statistics and The World Bank, https://data.worldbank.org/.

**Results**

Table 2 presents the parameter estimates of our electricity load specifications. All three models provide an impressive fit: their $R^2$s are all higher than 0.92. In the first column we report our base model (1). All parameters have the anticipated signs and magnitudes. The temperature parameters estimate an asymmetric V-shaped relationship, with one degree higher than the threshold of 62°F increasing electricity load roughly as much as three degrees below. A possible explanation for this feature is that, in Italy, a significant share of heating is generated by natural gas. Day-of-the-week and public-holidays dummy variables indicate that consumption drops significantly during weekends and festivities. It is worth mentioning that directly interpreting these coefficients as semi-elasticities is incorrect since the derivative of a function with respect to a dichotomous variable does not exist. Nevertheless, semi-elasticities can be computed by a simple rescaling operation, which depends on both the sign and the magnitude of the coefficient (e.g. Halvorsen and Palmquist, 1980). For example, the coefficient of Sundays and public holidays is -0.237. The appropriate decrease in load compared to Mondays (the baseline category) is 21.1% and not 23.7%.[14]

[ Table 1 about here ]

The focus of our analysis are the parameters $\gamma^*_{w,2020}$, i.e. the coefficients of the interaction terms between year 2020 and the weekly dummy variables, reported in the rows from $week_{1,2020}$ to $week_{22,2020}$. The parameters of the first 10 weeks are all non-significant, indicating that until the beginning of March there were no unobserved factors distinguishing the pattern of electricity load in year 2020 from the weekly averages of the previous five years. Therefore, our specification passes the in-time placebo test. We interpret this result as reassuring on the causal interpretation of our estimates. More specifically, the impacts estimated by the parameters $\gamma^*_{w,2020}$ from the 11th week onwards can be traced back to the COVID-19 and not to any other unobserved factors. All these parameters are negative and strongly significant: consistently with our expectations, our model estimates that the containment policies caused significant reductions in electricity load. The strongest reductions happened during the three weeks of most intense restrictions (weeks 13 to 15), where we estimate a causal effect of about –20% on electricity load.

---

[14] Semi-elasticities $\xi$ can be obtained using the following rescaling equation: $\xi = 100(\exp\{c\} - 1)$, where $c$ indicates the coefficient of the dummy variable (Halvorsen and Palmquist, 1980). This equation assumes that the coefficients is known, while in reality it is a random variable. The correct formula depends on the distribution of the error term (Van Garderen and Shah, 2002) but, in most cases, makes little difference, particularly when the parameters are highly significant. We tested different approximations but none changed our results in a meaningful way and, therefore, we opted for the above equation for simplicity.

In the second column, we report the specification in which we removed temperature in order to examine if our results are affected by the omission of relevant explanatory variables. The coefficient of the 2$^{nd}$ week of year 2020 is now significant at the 5% level, signaling that this model does not capture some of the distinctive features of electricity consumption during the current year and, therefore, fails the in-time placebo test. Here the reason is obviously the omitted temperature: the 2$^{nd}$ week of 2020 was, in fact, significantly colder than the average of the previous five years and the corresponding dummy variable is capturing this discrepancy. Incidentally, this result confirms that our in-time placebo test is powerful enough to detect unobserved factors. Despite the temperature omission, the $\gamma^*_{w,2020}$ coefficients measuring the impact of COVID-19 (i.e. weeks 11-22) remain remarkably stable and, therefore, our results are essentially unaffected by this misspecification.

The third column reports the maximum likelihood estimates of the model with an AR(1) error term. In this specification the R$^2$ cannot be directly compared with those of the previous regressions, since this index does not take into account the parameter of random component. Instead, we can compare the Akaike's Information Criteria and the log-likelihood. Both values clearly indicate that the performance of Model 3 is the superior one. The $\phi$ coefficient is about 0.6, signaling substantial autocorrelation in the error component. The standard errors of this model are somewhat larger than those of the previous specification. This feature suggests that OLS (and also the HAC standard errors) slightly underestimate the uncertainty of our findings. Apart from this difference, all the parameters appear to be consistent with those of the simpler specifications.

[ Figure 3 about here ]

Figure 3 represents the dynamic impact of the lockdown and social-distancing measures according to our best-fitting specification (Model 3), calculated as in (2). In line with our parameter estimates, impacts are non-significant for all the weeks before the outbreak. Starting from the introduction of the first lockdown measures in the second week of March, each incremental constraint (once again, see Table A1 for the timeline of the lockdown policies in Italy) is followed by a significant drop in estimated consumption, $\hat{Y}_t$, with respect to the counterfactual $\hat{Y}_t^*$. In a similar fashion, the loosening of the restrictions, which started on the 3$^{rd}$ week of April, prompted a resumption of economic activities and, therefore, our estimated negative impact steadily diminishes.

Table 2 summarizes our results on a monthly basis by reporting COVID-19 impacts on both electricity consumption and GDP according to Model 3 (Model 1 provides similar results and it is therefore omitted to improve readability). We present GDP impacts calculated using the two approaches described by equations (3) and (4). In bold we indicate what we believe is the preferred method for during each month. While residential consumption increased considerably during the lockdown, this tendency is likely to have reversed in May, when the restrictions limiting people's movements were lifted. During the month of March our best estimates conclude that the pandemic caused a 15.4% drop in GDP which, rescaled to the entire first quarter, leads to a -5.1% overall impact, with the 95% confidence interval being from -6.2% to -4.0%. Our mean value is remarkably close to the recession of 5.3% projected by Eurostat (2020) and the OECD (2020). However, it is worth mentioning that the interpretation of our results is different from those of the official statistics. They estimate the variation of GDP from the previous quarter, while our model isolates, *ceteris paribus*, the impact of COVID-19. Nevertheless, in our case-study the two measures are not likely to differ substantially given the sluggish conditions of the Italian economic system before the outbreak.

[ Table 2 about here ]

Comparing results over time, the most affected month is April, with a GDP reduction of roughly 25%. On the other hand, we start to witness some signs of recovery in May, for which estimate an average impact of -11%. While by the end of May the hardest restrictions had already been lifted, social-distancing requirements still constrained many social and economic activities. For example, shops and restaurants were open with reduced capacity, while the tourism sector had not yet restarted (Italy opened to domestic and international tourism during the first week of June). Therefore, it is certainly too early to observe a full comeback of the economy, or even just a temporary rebound triggered by the rescheduling of the industrial processes halted by the three-weeks shutdown in the months of March and April. A further issue, which needs to be considered when interpreting our most recent estimates, is international spillover effects. Production (Backus and Kehoe, 1992) consumption (Cavaliere et al., 2008) and financial assets (Forbes and Rigobon, 2002) of one country do not exist in isolation but are inextricably connected to the economic conditions of the rest of the word. Since Italy was the first European country to introduce strict lockdown policies, we believe that our estimates do correspond to the short-run, causal impact of such first restrictions. However, by mid-April 2020 the crisis had spread across the whole world and, therefore, we believe it is more cautious to interpret our estimated impacts for the most recent weeks as the general consequences of the pandemic, rather than tie down the entirety of their effect to specific changes in national policies.

**Conclusions and caveats**

This paper developed a simple and yet powerful methodology for estimating the impact of COVID-19 on the economy by analyzing high-frequency electricity market data. The first advantage of our approach lies in its real-time nature. While official statistics (e.g. Eurostat, 2020; OECD, 2020) are typically published with a delay of about three months, our method provides updated estimates of the economic impact of the pandemic on a weekly (or even daily) basis. In the current uncertain economic environment, timeliness is of essence for policy makers to understand the current state of the economy. Our approach can be used to monitor in real-time the extent of the disruption caused by the pandemic. It can also be used to assess the effectiveness of the the monetary and fiscal stimuli that many countries have introduced in order to address the crisis. A further strength of our empirical strategy is that it is widely applicable. It only requires data on electricity load and temperature, and such information is publicly available for virtually all developed economies in the world.

We illustrated our methodology using daily information from the day-ahead Italian electricity market. Italy was the first European country to experience the outbreak of COVID-19 and to implement a series of nation-wide restrictions on social and economic activities in an attempt to curb the spread of the infection. Aggregating our weekly results (robust to different specifications and to in-time placebo testing) we estimate that the impact on the GDP during the $1^{st}$ quarter of 2020 was -5.1%, which is very close to the preliminary figures published by Eurostat (2020) and the OECD (2020). Not surprisingly, we estimate that the most severe impacts were registered during the three weeks of March and April when all non-necessary economic activities were required to halt production. During this period, our preferred specification estimates a 30% GDP reduction. In the following weeks we detected a gradual recovery and, at the end of May, the level of economic activity appears to be roughly 7% lower than what it would have been if the pandemic had never happened. This difference is not surprising since social-distancing requirements are still constraining social and economic activities, particularly in the cultural, catering and hospitality sectors. Our model can be routinely updated over time to monitor GDP dynamics and to significantly anticipate the official statistics.

However, several caveats need to be considered when interpreting our findings. First, while we believe that at the national level postulating a 1:1 relationship between electricity load (excluding residential users) change and GDP change is satisfactory in the short-run, our approach does not disentangle the heterogeneous impacts of the pandemic across sectors. Some activities, such as food wholesale, retail and the e-commerce, are likely to have experienced an increase in turnover, while the small and medium companies in the manufacturing, hospitality and personal services have been the most vulnerable enterprises.[15]

Second, our analysis concerns only short-run effects. In fact, technological change can weaken the relationship between electricity load and GDP, making our assumption less accurate in the long-run. Additionally, with time passing by, the impacts of the different national policies tend to overlap with each other, and spillover effects from other countries become gradually more important. Economic shocks are not limited by frontiers but, rather, reverberate across countries almost like a different form of contagion, which consequences for both real (e.g. Backus and Kehoe, 1992) and financial markets (Forbes and Rigobon, 2002). This issue is likely to be negligible for our estimated GDP impacts of the initial weeks of restrictions, since Italy was the first country in Europe to grapple with the pandemic and implement lockdown policies. However, our results for the most recent weeks should be interpreted as the general effect that the pandemic, rather than only the response to a specific national policy. Despite these confounding factors and knock-on effects, the validity of our approach for monitoring the real-time status of the economy still stands.

Finally, our approach does not take into account that certain activities can be executed in smart-working. In this respect, while our results should be informative on the economic impact of further lockdown measures in the eventuality of a new outbreak, companies are now more prepared and economic disruption could be less severe. On the other hand, months of restrictions have put under enormous strain the whole economic system, and a second outbreak could have even more serious consequences if not contained efficiently.

---

[15] OECD statistics: http://www.oecd.org/sdd/business-stats/statistical-insights-small-medium-and-vulnerable.htm

Fezzi, C., and Bunn, D. (2010) Structural analysis of electricity demand and supply interactions. *Oxford bulletin of economics and statistics*, 72(6), 827-856.

Fezzi, C., and Mosetti, L. (2020) Size matters: Estimation sample length and electricity price forecasting accuracy. *The Energy Journal*, 41(4).

Forbes, K. J., and Rigobon, R. (2002) No contagion, only interdependence: measuring stock market comovements, *The Journal of Finance*, 57(5), 2223-2261.

Galimberti, J., K. (2020) Forecasting GDP Growth from Outer Space. *Oxford Bulletin of Economics and Statistics. 0305–9049.*

Halvorsen, R., & Palmquist, R. (1980) *The interpretation of dummy variables in semilogarithmic equations. American economic review*, 70(3), 474-475.

Henderson, J. V., Storeygard, A., & Weil, D. N. (2012) Measuring economic growth from outer space. *American economic review*, 102(2), 994-1028.

Hothorn, T., Zeileis, A., Farebrother, R. W., Cummins, C., Millo, G., Mitchell, D., and Zeileis, M. A. (2019) Package 'lmtest'.

International Energy Agency (IEA) (2020) *Global Energy Review 2020*, IEA, Paris. Available at: https://www.iea.org/reports/global-energy-review-2020.

Jan van Garderen, K., & Shah, C. (2002) Exact interpretation of dummy variables in semilogarithmic equations. *The Econometrics Journal*, *5*(1), 149-159.

Kraft, J., & Kraft, A. (1978). On the relationship between energy and GNP. *The Journal of Energy and Development*, 3(2), 401-403.

Malinauskaitea J., Jouharab H., Ahmad L., Milanic M., Montorsic L., Venturelli M. (2019) Energy efficiency in industry: EU and national policies in Italy and the UK, *Energy*, 172, pp. 255-269.16

# Tables and Figures

**Table 1**: Electricity load equation estimates

|   | **Model 1** Base | | **Model 2** No temperature | | **Model 3** Autocorrelation | |
| --- | --- | --- | --- | --- | --- | --- |
| Intercept | 10.47*** | 0.01 | 10.37*** | 0.01 | 10.42*** | 0.02 |
| Tue | 4.00*** | 0.29 | 4.04*** | 0.31 | 4.04*** | 0.18 |
| Wed | 4.74*** | 0.29 | 4.83*** | 0.31 | 4.85*** | 0.23 |
| Thu | 4.59*** | 0.29 | 4.72*** | 0.31 | 4.74*** | 0.25 |
| Fri | 3.71*** | 0.29 | 3.71*** | 0.31 | 3.69*** | 0.25 |
| Sat | -11.91*** | 0.29 | -11.90*** | 0.31 | -11.83*** | 0.23 |
| Sun | -23.70*** | 0.29 | -23.68*** | 0.31 | -23.73*** | 0.18 |
| $d_{holiday1}$ | -21.34*** | 0.50 | -21.30*** | 0.53 | -18.67*** | 0.33 |
| $d_{holiday2}$ | -5.42*** | 1.04 | -5.47*** | 1.11 | -0.40 | 0.68 |
| Temp | -0.21*** | 0.03 | - | - | -0.08* | 0.03 |
| $(Temp-62)d_{62}$ | 0.83*** | 0.05 | - | - | 0.37*** | 0.06 |
| week $_{1,2020}$ | 0.97 | 1.44 | 1.40 | 1.54 | 1.76 | 2.32 |
| week $_{2,2020}$ | 2.13 | 1.44 | 3.01* | 1.54 | 3.00 | 2.46 |
| week $_{3,2020}$ | 2.32 | 1.44 | 2.28 | 1.54 | 2.48 | 2.48 |
| week $_{4,2020}$ | 2.09 | 1.44 | 1.60 | 1.54 | 1.15 | 2.48 |
| week $_{5,2020}$ | 0.93 | 1.44 | -0.06 | 1.54 | 0.49 | 2.48 |
| week $_{6,2020}$ | 0.31 | 1.44 | 0.06 | 1.54 | -0.22 | 2.48 |
| week $_{7,2020}$ | 0.11 | 1.44 | -0.47 | 1.54 | -0.18 | 2.48 |
| week $_{8,2020}$ | -0.25 | 1.44 | -0.86 | 1.54 | -0.79 | 2.48 |
| week $_{9,2020}$ | -1.19 | 1.44 | -1.60 | 1.54 | -1.53 | 2.48 |
| week $_{10,2020}$ | -0.01 | 1.44 | 0.11 | 1.54 | -2.58 | 2.48 |
| week $_{11,2020}$ | -8.15*** | 1.44 | -8.48*** | 1.54 | -9.47*** | 2.48 |
| week $_{12,2020}$ | -18.71*** | 1.44 | -18.36*** | 1.54 | -16.30*** | 2.48 |
| week $_{13,2020}$ | -23.84*** | 1.45 | -22.66*** | 1.54 | -21.93*** | 2.48 |
| week $_{14,2020}$ | -22.53*** | 1.44 | -21.62*** | 1.54 | -21.43*** | 2.48 |
| week $_{15,2020}$ | -25.58*** | 1.47 | -25.67*** | 1.56 | -24.22*** | 2.48 |
| week $_{16,2020}$ | -18.30*** | 1.44 | -18.35*** | 1.54 | -16.23*** | 2.48 |
| week $_{17,2020}$ | -10.82*** | 1.44 | -11.40*** | 1.54 | -14.33*** | 2.48 |
| week $_{18,2020}$ | -11.38*** | 1.44 | -12.04*** | 1.54 | -11.02*** | 2.48 |
| week $_{19,2020}$ | -10.85*** | 1.44 | -11.00*** | 1.54 | -11.38*** | 2.49 |
| week $_{20,2020}$ | -9.13*** | 1.45 | -7.34*** | 1.54 | -8.36*** | 2.49 |
| week $_{21,2020}$ | -7.30*** | 1.45 | -5.16*** | 1.54 | -6.92** | 2.53 |
| week $_{22,2020}$ | -5.13** | 1.67 | -7.37*** | 1.78 | -7.41** | 2.83 |
| weekly FE | YES | | YES | | YES | |
| $\phi$ | NO | | NO | | 0.67 | |
| $R^2$ | 0.940 | | 0.932 | | 0.926 | |
| Log-lik | 3884.95 | | 3753.55 | | 4259.99 | |
| AIC | -7597.89 | | -7339.09 | | -8345.97 | |

*Notes*: All parameters except the intercept multiplied by 100 to improve readability. Significance levels are * =0.05, ** = 0.01 and *** = 0.001. Model 1 and 2 estimated with OLS (HAC standard errors returned slightly smaller intervals and, therefore, conservative inference, while model 3 is estimated with maximum likelihood. All models include 52 weekly fixed effects. N = 1979.

**Table 2**: Estimated monthly impacts of COVID-19

|  | *Electricity (%)* | *GDP$_1$ (%)* | *GDP$_2$ (%)* |
|---|---|---|---|
| *March* | -10.6<br>[-8.3; -12.9] | -13.7<br>[-10.7; -16.6] | **-15.4**<br>**[-12.1; 18.8]** |
| *April* | -16.8<br>[-14.6; -18.9] | -21.7<br>[-18.8; -24.3] | **-24.5**<br>**[-21.2; 27.5]** |
| *May* | -8.6<br>[-6.1; -10.8] | **-11.1**<br>**[-7.9; -13.9]** | -12.5<br>[-8.9; 15.7] |

*Notes*: Results according to Model 3 in Table 1. Electricity indicates the estimated impact on electricity load, GDP$_1$ indicate the impact on the GDP assuming that residential consumption has not changed and GDP$_2$ indicates the impact on the GDP assuming that residential consumption increased by 40%. The square brackets report 95% confidence intervals calculated via 5000 Monte Carlo simulations. In bold we highlight our preferred GDP change estimates.

**Figure 1:** Daily electricity consumption in years 2019 and 2020

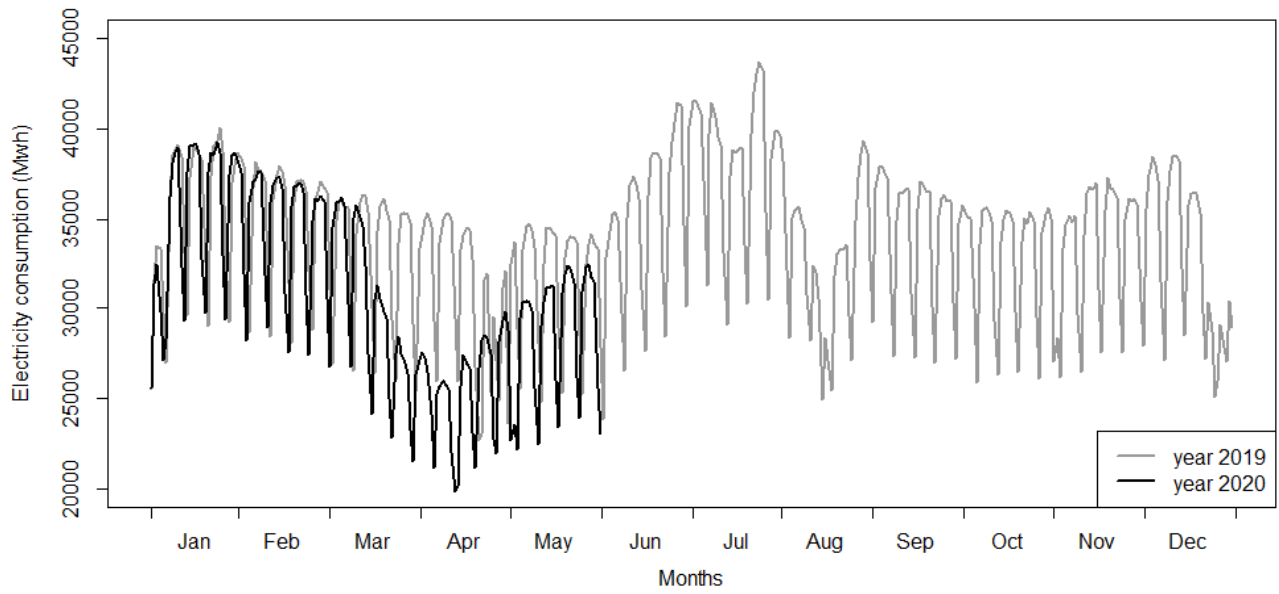

**Figure 2:** Average electricity consumption in each day of the week

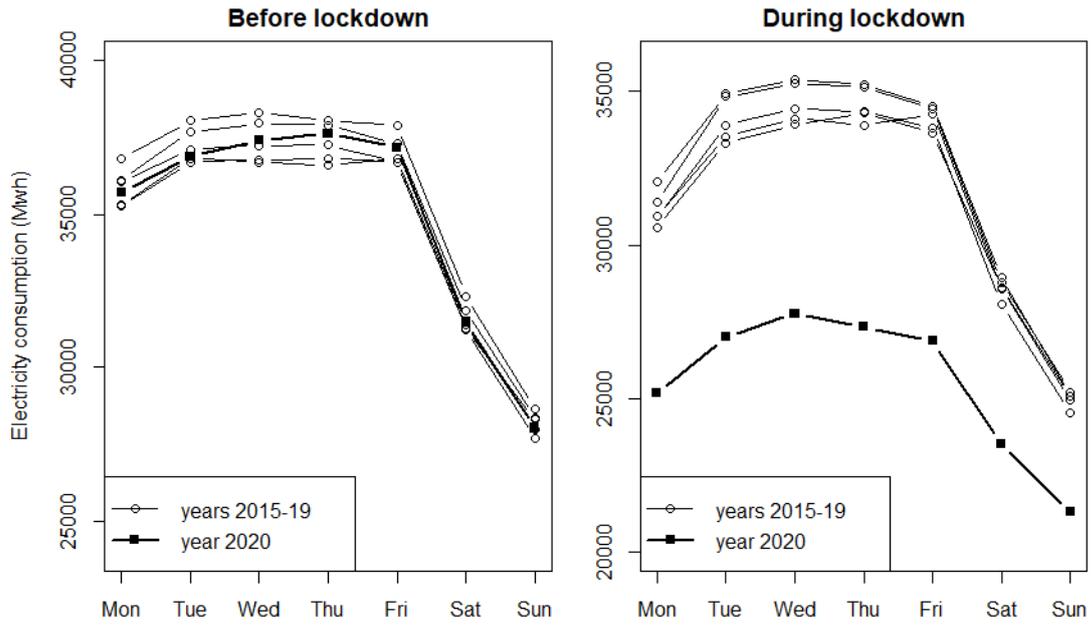

Notes: For all years the "before lockdown" period consists in weeks 5 to 9 (35 days in February and March), while the "during lockdown" period consists in weeks 12 to 16 (35 days in March and April).

**Figure 3:** Estimated impact of COVID-19 on electricity load

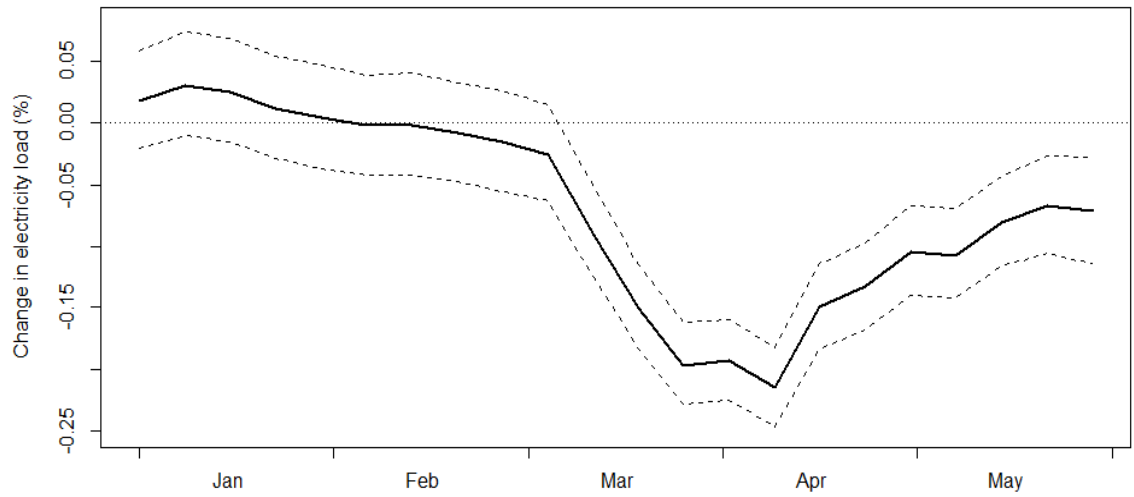

*Notes*: The solid line is the estimated impact of COVID-19 on electricity load according to our preferred specification, i.e. Model 3, while the dashed line representing the 95% confidence intervals calculated via 5000 Monte Carlo repetitions.

**Appendix: Supplementary Information**

**Table A1:** COVID-19-containment measures' timeline

| Date | Measures |
|---|---|
| *10/03/2020* | **Phase I begins**: the entire country is in lockdown |
| *12/03/2020* | *Lockdown tightening*: all shops (except supermarkets, pharmacies, stationers, gas stations), bars, restaurants, hair and nail salons are required to remain closed. |
| *14/03/2020* | Severe restrictions to domestic and international travels. Very significant reduction in transportation. |
| *23/03/2020* | *Lockdown tightening*: all unnecessary work activities a forced to shut-down. Only supermarkets, pharmacies, and other essential services are allowed to operate. |
| *14/04/2020* | *Lockdown loosening*: some activities, such as bookshops, manufacturing and major infrastructure projects are allowed to re-open. |
| *27/04/2020* | Some regions with a low number of COVID-19 cases allow certain commercial activities to re-open and lift restrictions on regional mobility. |
| *04/05/2020* | *Lockdown ending*: the entire country enters **Phase II**. Commercial and production activities that can enforce the social-distancing protocol can re-open. Bars and restaurants can operate only for take-away service. Hairdressers are still closed. Restrictions to mobility across regions are lifted. |
| *18/05/2020* | **Phase II.bis**: Most regions allow bars, restaurants and hairdressers to re-open following strict social distancing protocols. Regions can autonomously decide which activities are allowed to re-open. |

**Figure A1**: Scatter-plot of air temperature and electricity consumption

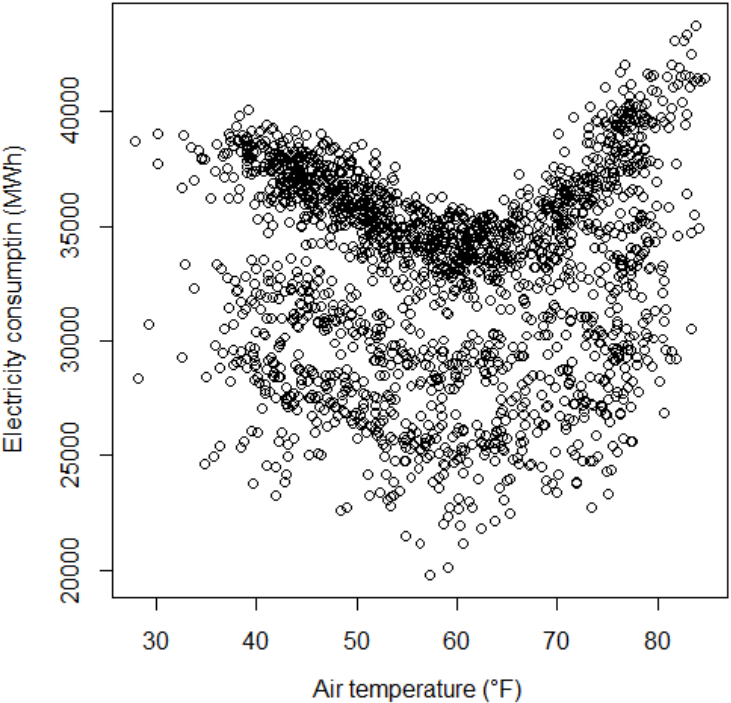

*Notes*: air temperature calculated as the daily average between Rome and Milan.